\documentclass[prd,nofootinbib,twocolumn,showpacs,superscriptaddress]{revtex4}

\usepackage{graphicx}
\usepackage[dvips]{color}
\usepackage{amssymb}

\begin{document}

\newcommand{\bA}{\mathbf{A}}
\newcommand{\bD}{\mathbf{D}}
\newcommand{\bE}{\mathbf{E}}
\newcommand{\bB}{\mathbf{B}}
\newcommand{\bX}{\mathbf{X}}

\newcommand{\VEV}{\eta}
\newcommand{\Lag}{\mathcal{L}}

\title{Abelian Higgs Cosmic Strings: Small Scale Structure and Loops}

\newcommand{\addressSussex}{Department of Physics \& Astronomy, University of Sussex, Brighton 
BN1 9QH, U.K.}

\author{Mark Hindmarsh}
\email{m.b.hindmarsh@sussex.ac.uk}
\affiliation{\addressSussex}

\author{Stephanie Stuckey}
\email{s.r.stuckey@sussex.ac.uk}
\affiliation{\addressSussex}

\author{Neil Bevis}
\email{n.bevis@imperial.ac.uk}
\affiliation{Theoretical Physics, Blackett Laboratory, Imperial
  College, London, SW7 2BZ, United Kingdom}

\begin{abstract}  
{
Classical lattice simulations of the  Abelian Higgs model are used 
to investigate small scale structure and loop distributions in 
cosmic string networks. 
Use of the field theory ensures that the small-scale physics is captured correctly. The results confirm
analytic predictions of Polchinski \& Rocha \cite{Polchinski:2006ee}
for the two-point correlation function of the string 
tangent vector, with a power law from length scales of order the string core width up to horizon scale with 
evidence to suggest that the small scale structure builds up from small scales.
An analysis of the size distribution of string loops gives a very 
low number density, of order 1 per horizon volume, 
in contrast with Nambu-Goto simulations. 
Further, our loop distribution function does not  support the detailed analytic predictions for loop 
production derived by
Dubath et al.\ \cite{Dubath:2007mf}.  Better agreement to our data is found with a model based on loop 
fragmentation \cite{Scherrer:1989ha}, coupled with a constant rate of energy loss into massive 
radiation. 
Our results show a strong energy loss mechanism which allows the string network to scale without gravitational radiation,  but which is not due to the production of string width loops. From evidence of small scale structure we argue a partial explanation for the scale separation problem of how energy in the very low frequency modes 
of the string network is transformed into the very high frequency modes of gauge 
and Higgs radiation. We propose a  picture of string network evolution which  reconciles the apparent 
differences between Nambu-Goto and field theory simulations. 
}
\end{abstract}


\maketitle


\section{Introduction}
\label{intro}

Cosmic strings are line-like objects formed in the early universe (for reviews see
\cite{0034-4885-58-5-001, vilenkin, Sakellariadou:2006qs}). 
They exist as solitons in theories with spontaneously broken symmetries if the vacuum manifold is not
simply connected \cite{Kibble:1976sj}, or 
as fundamental objects in string theory \cite{Copeland:2003bj}.
A network of cosmic strings may form in thermal phase transitions \cite{Kibble:1976sj}, 
at the end of hybrid inflation \cite{Yokoyama:1989pa,Kofman:1994rk,Copeland:1994vg}, or 
at the end of brane inflation when brane and anti-brane annihilate \cite{Sarangi:2002yt,Dvali:2003zj}.
They remain of immense cosmological interest through their appearance in these high
energy physics models, and further motivation is provided from the
enhanced fit of the $\Lambda$CDM model to Cosmic Microwave Background (CMB) data 
when cosmic strings are included \cite{Battye:2006pk, Bevis:2007gh} (see also \cite{Bevis:2004wk,Wyman:2005tu, Fraisse:2006xc, Battye:2007si, Urrestilla:2007sf}). Calculations of the CMB signal at small angular scales \cite{Fraisse:2007nu, Pogosian:2008am} and in the polarisation B-mode \cite{Bevis:2007qz,Pogosian:2007gi} show that 
future CMB observations will further constrain cosmological models with strings.

Cosmic strings form networks of infinitely long string
and loops, where a string can be called ``infinite'' in cosmological terms if it is
larger than the horizon. Infinite string takes the form of a random walk with 
correlation length $\sim \xi$. The network evolves in a self-similar manner, keeping $\xi$ at about the 
horizon scale, 
an important dynamical feature known as \textit{scaling}.
Scaling means that the energy density of infinite strings decreases 
as $1/t^2$, ($t$ is cosmic time), and thus constitutes a constant fraction of the total.

Since the original string network scaling paradigm was introduced 
\cite{Kibble:1976sj,Kibble:1980mv,Vilenkin:1981kz} 
a broad picture of the cosmological evolution of string networks has emerged, 
using a mixture of calculation, numerical simulation, and analytic modelling. 
Yet, a major unsolved problem is the eventual destination of the energy in the infinite strings.  This is of 
notable importance and greatly limits our ability to constrain string scenarios via their decay products, 
such as gravitational waves or energetic particles. 

In the traditional picture, string will loop back on itself and undergo a self-intersection event, which then 
results in the loop being cut from the long string. 
Strings have tension, so loops tend to collapse inwards and can shrink to a point. In their final moments, 
as their radius becomes close to the string width, they give up their energy in a burst of particle emission, 
which in principle might be detectable via cosmic rays. However whilst they are shrinking, they would 
oscillate and emit gravitational waves.
Conventionally it is taken that the loops are large relative to the
string width and that these gravitational waves take away most of the
energy.

Numerical simulations using the Nambu-Goto approximation
\cite{PhysRevD.40.973, Bennett:1989yp,Allen:1990tv}, in which the strings are modelled as relativistic 
one-dimensional entities, seemed to support this picture. Copious loop production was observed at scales a small fraction of the cosmological horizon, along with the presence of small-scale structure in the correlation functions of the long string. The small-scale structure is related to the creation of small-scale loops, but progress on understanding the connection has been slow.

It is crucial to establish the dominant length scale of loop production, as it controls both the amplitude and frequency of the gravitational wave signal, and the fraction of energy going into ultra-high energy cosmic rays.
The current debate is summarised in Section V. Here we merely note that there is a great deal of uncertainty over the loop population of cosmic string networks and the Minkowski-space Nambu-Goto simulations have been used to argue that loop production might even peak on the scale of the string width \cite{Vincent:1996rb}.

\begin{figure}
\scalebox{0.4}{\includegraphics{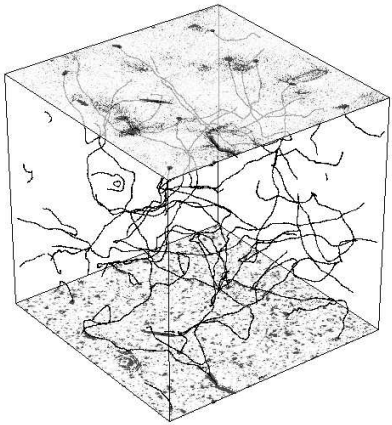}}
\caption{\label{radiating}
{A snapshot from a simulation of a string network in the Abelian Higgs model.
Lines show the centres of the strings, and the shading on the top and bottom faces 
represents magnetic field
  energy density and Higgs field potential energy density respectively.
   }
   }
\end{figure}

If this is true, 
it is necessary to 
go beyond the Nambu-Goto approximation and calculate with the underlying theory.
The Nambu-Goto approximation also breaks down at kinks (discontinuities in the string tangent vector) and cusps (points where the tangent vector vanishes and the string doubles back), which are universal and common features of a string network.
The underlying theory for solitonic strings is a 
quantum field theory: Fortunately, quantum corrections appear to be small \cite{Borsanyi:2007wm}, 
and classical field theory should be a good approximation.
Numerical simulations in the classical Abelian Higgs model \cite{PhysRevLett.80.2277,Moore:2001px} 
showed that infinite string does indeed scale by losing energy into 
gauge and Higgs radiation (see Fig.\ \ref{radiating}), although it was not established whether 
the decay proceeded via 
short-lived loops at the size of the string core width, or directly from the long strings themselves. 
In any case, no sign of copious loop production was found.
In this paper we present evidence that direct radiation is much more important than small loop production for Abelian Higgs string networks.

\begin{figure}
\scalebox{0.4}{\includegraphics{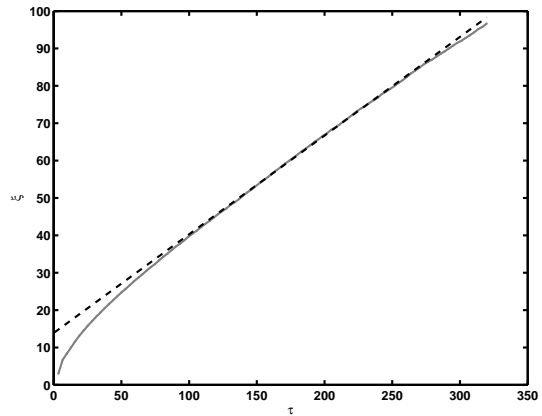}}
\caption{\label{bigScaling}
{The comoving network length scale $\xi$ (roughly corresponding to the radius of curvature of long 
string) for a $1024^3$ lattice simulation in the matter era. The linear behaviour with conformal time $\tau
$ is evidence for scaling extending to 
values of $\xi$ nearly two orders of magnitude greater than the string width, which is of order unity in simulation units. See Section \ref{LAHmodel} for more information.
}
}
\end{figure}

The idea of direct radiation raises a puzzle:  in order to create radiation in a mode with mass $M$, the field must be oscillating 
at a frequency $\omega \sim M$.  A smooth string curved on the horizon scale $H^{-1}$ is constructed 
from field modes with frequencies $\omega \sim H$.  In view 
of the mismatch it was argued that gauge and Higgs radiation must be negligible \cite{Everett:1981nj}.
Numerical simulations of smooth strings \cite{PhysRevLett.84.4288} show that 
radiation is indeed suppressed by a factor exponential in the ratio of the curvature radius to the string 
width. Nevertheless, simulations exhibit scaling behaviour at times when the network length scale 
exceeds the string width by a factor $\sim 50$ \cite{PhysRevLett.80.2277,Moore:2001px,Bevis:2006mj}, 
and for example in Fig.\ \ref{bigScaling} we see that this is confirmed to $\sim 100$ in our larger 
simulations.

It should be noted that a similar puzzle was presented by 
the small size (compared with the horizon scale) of loops in Nambu-Goto simulations. 
Recent work by Polchinski and collaborators 
\cite{Polchinski:2006ee,Polchinski:2007rg,Dubath:2007mf}
has resolved the problem by 
showing how small-scale structure can give rise both to apparently smooth strings 
and to loop production at the small-scale cut-off on the string network. 

Motivated by the success of the approach by Polchinksi and collaborators 
\cite{Polchinski:2006ee,Dubath:2007mf}, 
we analyse the small scale structure of 
cosmic strings in the Abelian Higgs model 
through the two-point
tangent vector correlator  
and the loop distribution function.
We find excellent agreement for the two-point tangent vector correlator, showing that its slope at short 
distances and the mean square velocity are related as predicted by their model but with mean square 
velocities  found to be significantly lower in Abelian Higgs simulations than in Nambu-Goto simulations.
The presence of small-scale structure on Abelian Higgs strings leads us to propose that the high frequencies required for direct radiation are generated in much the same way as small loops.

However, the prediction for the form of the loop production function is much less successful, even 
after taking into account the fact that in field theory simulations loops lose energy and shrink at a 
constant rate. From evidence found of fragmentation of horizon size loops we are prompted to compare 
our data to the model of Scherrer \& Press \cite{Scherrer:1989ha} and this seems to provide a better explanation.

We also investigate the speed with which the network relaxes to scaling, and in particular how quickly 
the small scale structure appears.  We find evidence that the relaxation time goes roughly as the initial network 
length scale. Furthermore, we present evidence that small scale structure grows from the bottom up, 
rather than resulting from a cascade from large scales as has hitherto been assumed 
\cite{Polchinski:2006ee}.

Our paper is laid out as follows.
A review of the traditional picture of energy loss from cosmic strings is
provided in Sec.\ \ref{CSD} then the particulars of the field theory
simulation for Abelian Higgs strings are outlined in
Sec.\ \ref{LAHmodel}. Results of the analysis of the two-point correlation
functions and loop distributions are discussed in
Sec.\ \ref{correlations} and Sec.\ \ref{loopDistributions}. We summarise our results in Sec.\ \ref
{summary} and present a picture of string network evolution which reconciles our results with 
those derived from Nambu-Goto simulations.


\section{{Energy Loss Mechanisms}}
\label{CSD}

Conventional ``solitonic'' cosmic strings are produced in a symmetry-breaking phase transition at a scale 
$\eta$, in a field theory with coupling constant $g$, and typical mass scale $m \sim g\eta$. 
They have linear mass density $\mu \sim \eta^2$ and have width $\delta \sim m^{-1}$. For example, 
strings produced 
at the Grand Unification energy scale have $\eta \sim 10^{16}$GeV, and gravitational coupling $G\mu  
\sim  (\eta/m_{pl})^2 \sim 10^{-6}$.

A particularly important feature in string energy loss is the presence of 
cusps, points where the speed of the string becomes the speed of light.
This produces a singularity in the shape of the string where the tangent vector to the loop vanishes. The 
string is bent back on itself and has the opportunity to interact with itself over the length of the cusp 
region. 
Also important are kinks, formed during intercommutation when there are sharp changes
in the tangent vector. 
These discontinuities then resolve themselves into kinked waves travelling in both directions away
from the intercommutation site. 
Numerical simulations based on the Nambu-Goto equations 
show a large number of intercommutations, forcing a build up of small scale
structure.
When oppositely-moving modes interact they can emit
 gravitational radiation, and the resulting back reaction 
\cite{Hindmarsh:1990xi} can smooth the kinks, as shown
in numerical simulations \cite{Quashnock:1990wv}.

Conventionally, three main energy loss mechanisms have been considered, all of which take place on 
loops which have broken off from the long string network. Perturbative production of particles from the 
coupling to the Higgs field for strings with $m \ll \mu^{1/2}$ 
can be calculated to produce a flux of high energy particles with power \cite{Srednicki}
\begin{equation}
\label{power_pp}
P_{p} = \mu \delta l^{-1},
\end{equation}
where $l$ is the size of the loop.
Where string is overlapping in cusp regions, it can annihilate releasing energy in the form of high energy 
particles. The original estimate \cite{Brandenberger} $P_{c} \propto \mu \delta^{1/3}l^{-1/3}$ was 
corrected 
in \cite{BlancoPillado:1998bv} who obtained 
\begin{equation}
\label{power_cusp}
P_{c} \propto \mu \delta^{1/2}l^{-1/2}.
\end{equation}
The power emitted through gravitational radiation, $P_{g}$, from a
sizeable oscillating string loop of length $l$ and mass $m \sim
\mu l$ can be estimated from its quadrupole moment $I \sim m l^2$ 
\cite{Weinberg, Vachaspati:1984gt}.
\begin{equation}
\label{power_grav}
P_{g} \sim G \Big(\frac{d^3 I}{dt^3}\Big)^2 \sim G\omega^6 I^2 \propto G\mu^2. 
\end{equation}

Much additional work has been done on the production of gravitational wave bursts at the sites of cusps 
on strings with and without additional small scale structure 
\cite{Damour:2000wa,Damour:2001bk,Siemens:2001dx,siemens:085017}.
Cusps on strings with small-scale structure are also sources of intense loop production 
\cite{Siemens:2003ra,Dubath:2007mf}.

Purely based on the length dependence of these relations for the power
output; gravitational radiation $P_{g}$  is the dominant decay
channel for loops of length $l > \delta (G\mu )^{-2}$, where $\delta$ is
the string core width. Below this length, cusp annihilation is 
is dominant. So it is of considerable importance to discover the distribution of different sized loops 
produced and evolving in the network as their size determines decay modes and the radiative by-
products. This has crucial bearing on estimates of gravitational radiation or flux from high energy cosmic 
rays that we may be able to detect from a network of cosmic strings.  

In the conventional cosmic string scenario the loop distribution is measured in simulations using the 
Nambu-Goto approximation and neglecting gravitational back-reaction.  This is justifiable when the 
radius of curvature of the string is much greater than the string
core width so a string can be assumed infinitely thin,
and when the weak field approximation holds, $G\mu \ll 1$. 
It is clear that in the presence of kinks and cusps, the Nambu-Goto approximation is strictly not justified, 
but it is assumed that gravitational radiation back-reaction will act to smooth the strings on a scale $l \sim 
(G\mu)^{1+2\chi}t$ \cite{Polchinski:2007rg}, where $\chi$ is a small parameter defined below, and that 
large-scale properties will be reproduced correctly.  

In the absence of back-reaction, Nambu-Goto simulations show that loops are produced at a small 
constant physical scale, which is most likely the initial correlation length of the network 
\cite{Ringeval:2005kr,martins:043515,Olum:2006ix}. In Ref.\ \cite{martins:043515} there is a claim that there are signs 
that this scale is growing, while Ref.\ \cite{Olum:2006ix} emphasises the significance of an apparently 
stable population of loops with sizes $l \sim 0.1t$, arguing that the peak at the initial correlation length 
will eventually disappear.  

In view of the automatic breakdown of the Nambu-Goto approximation when strings intercommute and 
produce kinks, it would seem sensible to simulate strings in an underlying field theory. One resolves the 
kinks and cusps correctly, and includes classical radiation as a form of energy loss.  While the 
conventional arguments given above emphasise gravitational radiation, omitted from all simulations, 
one should be able to see the other forms of energy loss and to check their scaling with loop size.  
However, as outlined in the introduction, previous field theory simulations 
\cite{PhysRevLett.80.2277,Moore:2001px} found a scaling infinite string network without 
a significant population of loops. There appears to be an energy-loss mechanism allowing the strings to 
scale, which has a different length dependence to any of those outlined above.

To see this, consider a string network whose spatial distribution scales with the horizon size such that the length of string in an horizon volume is $\sim t$. The energy density of the string network hence varies as $\rho_s \sim \mu/t^2$, which then requires the network to lose energy at a rate $\left| \dot{\rho_s}\right| \sim \mu/t^3$. Hence the rate of energy loss from a piece of string of \mbox{size $\sim t$} is $P_s \sim \mu$. Thus a scaling network requires an energy loss mechanism which is independent of the size of the string, like gravitational radiation, but a factor $(G\mu)^{-1}$ stronger.  One implication of the mechanism is that the length of a loop of string will shrink at a constant rate of order unity.

\begin{figure}
\includegraphics[width=0.9\columnwidth]{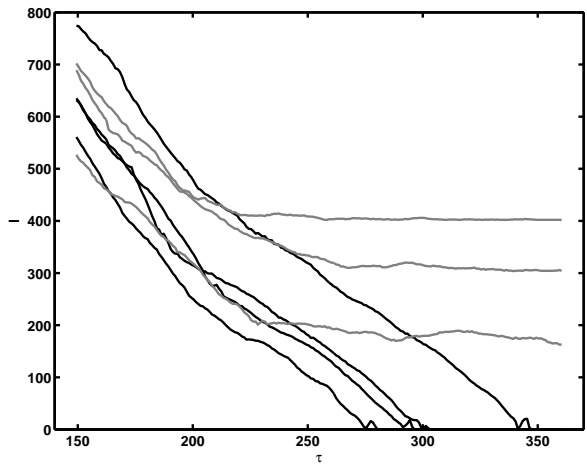}
\caption{\label{shrinking} Change of total comoving length of string $l$ with conformal time $\tau$ for  late time radiation era simulations of box side size $L=256\Delta x$ with $\Delta x=0.5$. The energy loss  from a  loop of string is linear with constant gradient until the loop fragments and evaporates (examples shown in black). Once only 2 straight strings remain in the box (3 examples shown in grey), the linear energy loss stops and each string length oscillates around the box crossing length; $L$, $L\sqrt{2}$ or  $L\sqrt{3}$ depending on the direction the string wraps the box. }
\end{figure}

We verify this by running $256^3$ radiation era simulations for a very long time until there is only a single shrinking loop or a pair of straight strings winding around one or more directions in the simulation volume, which has periodic boundary conditions. The change in total comoving string length with conformal time is shown in Fig.\ \ref{shrinking} to be linear. By inspection one can see that for strings that are shrinking, $d{l}/d\tau \sim \mathcal{O}(1)$ as claimed.
Once two strings in the box  become straight their length oscillates around the box crossing distance, which is the behaviour expected in the Nambu-Goto approximation.

The detailed mechanism for this strong energy loss is not well understood. 
Attention has been focused on the production of `core' sized loops 
\cite{PhysRevLett.80.2277,Moore:2001px}, which would nicely connect the Nambu-Goto and field theory simulations. Core loops 
produced {at the size of the string core width} would {quickly evaporate into classical radiation, and 
despite a large production rate
would have a very low number density, which is easily estimated to be a few per Hubble volume
 \cite{Moore:2001px}.}  In this scenario, field theory strings are behaving like Nambu-Goto strings in that 
loops are being produced at the smallest physical scale, which is in one case the string width, and the 
other the initial correlation length.
While we see core loops, we would expect them to be associated with long string, and as we will show, 
we are unable to find such a correlation.  Their number density is also too low to account for the energy loss from long string. It may be that energy is being broken off in lumps which are too 
small to register as loops at all.


\section{String simulation method}
\label{LAHmodel}

\subsection{Abelian Higgs model}

As discussed in the introduction, we characterise the small-scale structure and loop distribution of 
strings via simulation of the Abelian Higgs model 
\cite{0034-4885-58-5-001, vilenkin, Sakellariadou:2006qs}. This is the simplest field theory to contain local (or gauge) cosmic strings and has Lagrangian 
density: 
\begin{equation}
\label{Lagrangian}
\Lag = (D_{\mu}\phi)^{*}(D^{\mu}\phi) - \frac{1}{4e^{2}}F_{\mu\nu}F^{\mu\nu} - \frac{\lambda}{4}(|\phi|^{2}-
\VEV^{2})^{2}, 
\end{equation}
where $\phi$ is a complex scalar Higgs field, $D_{\mu} = \partial_{\mu} + iA_{\mu}$ is the gauge-
covariant derivative and $F_{\mu\nu}= \partial_{\mu}A_{\nu} - \partial_{\nu}A_{\mu}$ is the usual field 
strength tensor. String solutions are well known in this model \cite{Nielsen:1973cs}; the phase of the 
Higgs field winds around $2\pi n$ ($n \in \mathbb{Z}^{\pm}$) as a closed loop is traversed through 
space and $\phi$ is forced to depart from the vacuum manifold over a tube of \mbox{radius $\sim1/\sqrt
{\lambda}\VEV$}. The gauge field acts to compensate the winding but this results in a pseudo-magnetic 
flux tube  of \mbox{radius $\sim1/e\VEV$}. 

\subsection{String width control}

These two fixed physical length scales must be resolved in any simulation of the string network but, in an 
expanding Friedman-Robertson-Walker universe, they rapidly fall away from the other length scales that 
must also be resolved. Using comoving coordinates the horizon size is given simply by the conformal 
time $\tau$ and is the relevant length scale at which the super-horizon tangle of string begins to 
straighten and decay. However, the comoving string width decays as the reciprocal of the cosmic scale 
factor $a$ and therefore as $\tau^{-1}$ in a radiation-dominated universe and $\tau^{-2}$ under matter-
domination. We therefore require to resolve two scales which diverge as $\tau^{3}$ under matter 
domination, which would, in principle, limit us to very short periods of simulation\footnote{That our 
simulations resolve the string width limits them in size to being far smaller than the horizon size at 
matter-radiation equality and therefore our simulations are also limited to much earlier times. However, 
we can simulate a matter dominated universe at very early times and use scaling to make statements 
about a matter dominated era at later times.}.  

However in Ref. \cite{Bevis:2006mj}, a technique was demonstrated via which the coupling constants $e
$ and $\lambda$ were raised to time dependent variables:
\begin{eqnarray}
\lambda & = & \lambda_0 a^{-2(1-\kappa)}\\
e       & = & e_0 a^{-(1-\kappa)}
\end{eqnarray}
in order that the comoving string width $r$ behaves as:
\begin{equation}
r \propto a^{-\kappa}.
\end{equation}
That is, $\kappa=1$ gives the true dynamics while $\kappa=0$ yields a comoving string width, which is 
particularly convenient for simulation. The dynamical equations derived upon variation of the action 
were then:
\begin{eqnarray}
\label{e:EOMHiggs}
\ddot{\phi} + 2 \frac{\dot{a}}{a}\dot{\phi} - D_j D_j \phi 
= 
- a^{2\kappa} \frac{\lambda_0}{2}(|\phi|^{2}-\eta^{2})\phi \\
\label{e:EOMGauge}
\dot{F}_{0j} +2(1-\kappa)\frac{\dot{a}}{a}F_{0j}- \partial_{i}F_{ij} 
= 
-2a^{2\kappa} e_0^2 \mathcal{I}m[\phi^{*}D_{j}\phi]
\end {eqnarray}
(in the gauge $A_{0}=0$) with the $A_{0}$ variation yielding the quasi-Gauss' Law constraint:
\begin{equation}
-\partial_i F_{0i}
=
-2a^{2\kappa}e_0^2 Im[\phi^{*}\dot{\phi}].
\end{equation}
Although these field equations do not conserve energy if $\kappa\ne1$ (because the action is not time-
translation invariant), the dynamics were shown in Ref. \cite{Bevis:2006mj} to be insensitive to $\kappa$. 
Indeed the difference in their results for the two-point correlation functions of the energy-momentum 
tensor between $\kappa=1$ and $\kappa=0$ in the radiation era, where $\kappa=1$ was achievable, 
were found to be slight while their results for the CMB power spectra, which are dominated by the matter 
era, were found to be similarly insensitive to $\kappa$ over the range $0$ to $0.3$, with the later being 
the largest value at which reliable CMB results could be obtained.

Here we use the equations of motion Eqs.\ (\ref{e:EOMHiggs}) and (\ref{e:EOMGauge}) with $\kappa=0$ 
throughout, although we also check our results using simulations with $\kappa=1$ for the radiation era 
(only), and find changes that are negligible. Note, however, that our equations of motion at $\kappa=0$ 
are not precisely the same as those used in Ref. \cite{Moore:2001px}.

\subsection{Simulation Specifics}
\label{SimSpecifics}
Eqs.\ (\ref{e:EOMHiggs}) and (\ref{e:EOMGauge}) are discretised onto a lattice using the procedure 
described 
in Ref. \cite{Bevis:2006mj} and hereafter called LAH, which is an extension from Minkowski space-time 
to flat FRW universes of the standard approach of Ref. \cite{Moriarty:1988fx}. This preserves the Gauss' 
law constraint to machine precision. Initial conditions were chosen following Ref. \cite{Bevis:2006mj} 
in order for a string network to emerge rapidly and the Gauss constraint obeyed. To achieve the latter,  we set to zero the gauge field and the time derivatives of both the gauge and Higgs fields. To achieve the former, we set the simulation start time such that the horizon is comparable to the (uniform) lattice spacing $\Delta x
$ and therefore the phase of scalar field is an independent random variable on each lattice site, while 
we set $|\phi|=\VEV$. We employ periodic boundary conditions and therefore the fields can be  evolved
forward reliably until the half-box crossing time for light. 

We use a lattice spacing of $\Delta x=0.5$ and set $\VEV=1$, $\lambda_0=2$ and $e_0=1$, which 
together guarantee that strings are resolved for all $a$ when $\kappa=0$ and for $a\lesssim1$ in our 
radiation era $\kappa=1$ check. Note that the above scalings leave the ratio $\lambda/2 e^{2}$ constant 
and we therefore study the model at the Bogomol'nyi value \cite{Bogomolnyi:1976}, which yields equal 
scalar and gauge masses. The simulations were performed using the UK National Cosmology 
supercomputer \cite{Cosmos}, with parallel processing enabled via the LATfield library \cite{LATfield}, 
using a lattice size of $512^3$ and averaging over 20 realisations. Additional simulations using a $768^3$ lattice were performed on the University of Sussex HPC Archimedes cluster.

We have been able to access a dynamic range of similar order to Nambu-Goto simulations performed by other groups working on small scale structure issues but due to the differences between the simulations the measures for dynamic range that are suitable in one case are ambiguous in another, making this a difficult comparison. On $N^3$ lattice volumes
with $\Delta x=0.5$, a network of strings of width $\delta \sim 1$ is fully formed from conformal time $\tau_i \sim
20$ and is simulated until  $\tau_f = N\Delta x/2$, when 
the periodic boundary conditions of the simulation volume can
potentially be felt. Checks of the full energy-momentum tensor indicate that scaling is achieved at around $\tau_{sc} \sim 64$. 

One measure of dynamic range is  
$\xi(\tau_f)/\delta \sim 50$, which contains measurable quantities in our simulation. This can be  compared with 
the ratio of the initial and final correlation lengths, ($\sim 100$ \cite{Ringeval:2005kr,Olum:2006ix})  in Nambu-Goto simulations although the initial correlation length has no straightforward meaning in our simulation. Another measure is the ratio of the final time to the time at which the network achieves scaling. For us,  $\tau_{sc} \sim 64$ gives us  a dynamic range of about 2 (in conformal time) for our $512^3$ simulations and about 3 for $768^3$ lattices.   Nambu-Goto simulations \cite{Ringeval:2005kr} estimate a dynamical range of 5 from the scaling of the energy density of long string in the radiation era.\subsection{String length measurements}
\label{lengthMeasures}

\begin{figure}
\includegraphics[width=0.9\columnwidth]{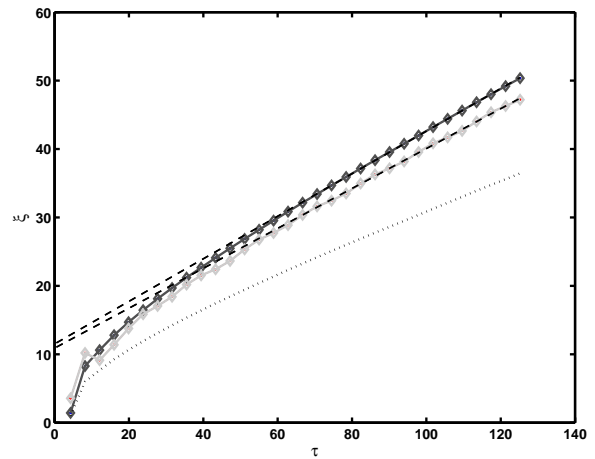}
\caption{\label{xiCompared}Network length scale $\xi$ in a radiation era simulation calculated from Eq.\ 
(\ref{xiDef}) is in solid dark grey with fit over scaling times ($\tau\in[64, 128]$) giving $\xi \propto 0.31\tau
$ (the dotted line shows $\xi$ before rescaling total string length using the Scherrer-Vilenkin factor). $
\xi_{\mathcal{L}}$ calculated directly from the Lagrangian is in solid light grey with fit $\xi_{\mathcal{L}} 
\propto 0.29\tau$.}
\end{figure}

At intervals during the evolution we record the coordinates of lattice plaquettes around which the phase 
of $\phi$ has a net winding number\footnote{While a winding of the phase is gauge-invariant in the 
continuum, on a lattice it can be removed by a finite gauge transform. Therefore we employ the gauge-
invariant measure of Ref. \cite{Kajantie:1998bg}}. As a first approximation we then take it that a segment 
of string having length $\Delta x$ threads through each plaquette of $2\pi$ winding and joins the centres 
of the lattice cells on either side. From these segments we can then construct the path of the string, 
although since it is composed of an array of perpendicular lengths the string length is overestimated. We 
do not attempt to smooth the paths in order to compensate, as in Refs. 
\cite{PhysRevLett.80.2277, Moore:2001px, Bevis:2008hg}, but instead apply a Scherrer-Vilenkin correction of $\pi/6$ 
\cite{Scherrer:1997sq} to such length measurements. This factor is derived from the length of the line $\phi=0$ for a two-component Gaussian random field $\phi$ and so will not be completely accurate for our dynamic string 
network. However, the results for the average string length density are in approximate agreement with 
that measured using the Lagrangian density Fig.\ \ref{xiCompared}. This second method makes use of 
the fact that perturbative radiation makes zero contribution to $\Lag$, while a static straight string 
contributes at $-\mu$ per unit length density. Since $\Lag$ is a four-scalar and length density picks up a 
$\gamma$-factor upon a Lorentz transform, then $-\int\Lag d^3x/\mu$ is a measure of the (invariant)
string length.

Rather than use the (comoving) string length density directly, we instead compare with the network {length 
scale} $\xi$, defined as:
\begin{equation}
\label{xiDef}
\xi = \sqrt{\frac{V}{L}},
\end{equation}
where $V$ is the reference volume and $L$ the string length within it. For a scaling network $\xi\propto
\tau$.  

\begin{figure}
\includegraphics[width=0.9\columnwidth]{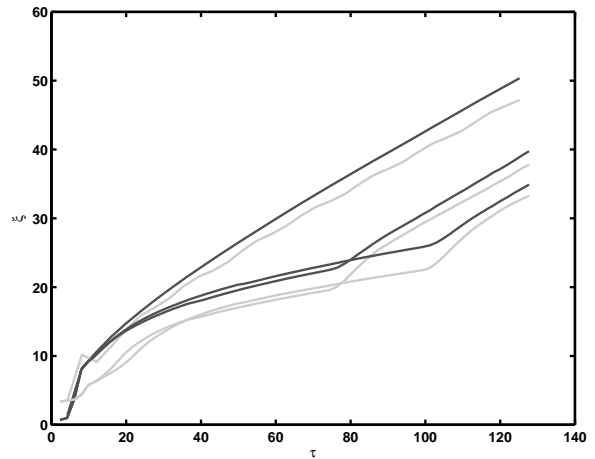}
\caption{\label{xi_scaling}Network length scale $\xi$, showing linear behaviour with conformal time 
once the 
  network settles into a scaling regime and memory of initial conditions
  is lost. $\xi$ (black) and $\xi_{\Lag}$ (grey) are shown for radiation era simulations according to Eqs.\ 
(\ref{e:EOMHiggs}) and (\ref{e:EOMGauge}) with parameters as
  stated in Sec.\ \ref{SimSpecifics}. 
  Curves for smaller $\xi$ from simulations with a constant
  damping term which is not switched off until $\tau=75$ and
  $\tau=100$ respectively showing the speed with which the network length scale
  resumes the same scaling evolution with $\xi\propto 0.31\tau$. 
}
\end{figure}

In Fig.\ \ref{xiCompared}, we plot both the Lagrangian measure result $\xi_{\Lag}$ and the winding 
measure result $\xi$ (with no subscript since it is our default measure), which reveal a linear behaviour $
\xi\propto(\tau-\tau_{\xi=0})$ after an initial transient, consistent with the expected scaling.  As pointed out in Ref. \cite{Bevis:2006mj}, there is nothing fundamental about the value of $\tau_{\xi=0}$ 
and it is simply an artefact of the initial conditions. Indeed, this can be seen in Fig.\ \ref{xi_scaling}, which 
shows additional results from two runs in which an artificial damping phase (similar to that used in Ref. 
\cite{PhysRevD.55.573}) was employed for an extended time. As can be seen in the figure, when the 
damping is released, both these runs show $\xi$ rapidly reverting to approximately the same gradient as 
the undamped run. 
Reference \cite{Bevis:2006mj} also observed scaling with $\tau-\tau_{\xi=0}$ in the two-point correlation 
functions of the energy-momentum tensor, so there is no evidence from the simulations that this scaling 
is a transient. These energy-momentum correlators show that the 
network demonstrates scaling in a $512^3$ simulation over the conformal time range $64<\tau<128$, 
which will be referred to as the scaling epoch.

\section{Correlation Functions}
\label{correlations}

\subsection{Tangent Vector Correlators}

A 2-point correlation function from LAH simulation data is compared to
analytic results which are reviewed in a
manifest form with a view to introducing the variables and functions under discussion. For a thorough analysis refer to
\cite{Polchinski:2006ee,Dubath:2007mf,Polchinski:2007rg}.  The
equations of motion are reformulated in terms of the left and right
moving unit vectors $\textbf{p}$ and $\textbf{q}$ defined in terms of the
position vector $\textbf{x}(\sigma,\tau)$ where $\sigma$ is the string
coordinate.
\begin{eqnarray}
\label{pqdefs}
\textbf{p}(\sigma,\tau)=\dot{\textbf{x}}+\textstyle{\frac{1}{\epsilon}}\textbf{x}' \textrm{  and  }
 \textbf{q}(\sigma,\tau)=\dot{\textbf{x}}-\textstyle{\frac{1}{\epsilon}}\textbf{x}'
\end{eqnarray}
where dots are derivatives with respect to $\tau$, primes with respect to $\sigma$ and
\begin{equation}
\label{eqn4eps}
 \epsilon^2=\frac{{\textbf{x}^{\prime}}^2}{1-\dot{\textbf{x}}^2} 
\end{equation}
  thus giving the
 equations of motion in the transverse gauge \cite{PhysRevD.40.973}
\begin{eqnarray}
\label{EOM}
\dot{\textbf{p}} - \textstyle{\frac{1}{\epsilon}}\textbf{p}'&= - \textstyle{\frac{\dot{a}}{a}}( \textbf{q} - (\textbf{p}
\cdot \textbf{q})\textbf{p} ) \\
\dot{\textbf{q}} + \textstyle{\frac{1}{\epsilon}}\textbf{q}' &= - \textstyle{\frac{\dot{a}}{a}}(
\textbf{p} - (\textbf{p}\cdot\textbf{q})\textbf{q} ).
\end{eqnarray}
From the coordinate locations of the string extracted from LAH it is simple to
compare the Euclidean distance between 2 points on the string and the
separation along the string coordinate.

The longest string at a set of equally spaced times throughout
the scaling epoch in the simulation is isolated for analysis in 
both the radiation and matter dominated eras. The comoving distance
along the string $s=\int\epsilon \textrm{d}\sigma$ along a string coordinate length $\sigma =\sigma_1-
\sigma_2$ 
is
compared to the Euclidean distance, $r$, between $\sigma_1$ and
$\sigma_2$. Along each loop of string, the mean square Euclidean
distance is averaged over many starting
points a few string segments apart around the loop, thus creating a 2-point
function, 
\begin{equation}
\langle r^2(\sigma,\tau)\rangle = \int_0^{\sigma}\int_0^{\sigma} \textrm{d}^2\sigma \,\langle 
\textbf{x}'(\sigma_{1}) \cdot
 \textbf{x}'(\sigma_{2})\rangle.
 \end{equation}
 
For consistency with notation used elsewhere \cite{Polchinski:2006ee, martins:043515} and  using $
\textbf{x}' = \frac{\epsilon}{2} (\textbf{p} - \textbf{q})$ in terms of the
 definitions of Eq.\ (\ref{pqdefs})  we denote
\begin{eqnarray} 
\label{corrX}
\textrm{corr}_{\textrm{x}}(\sigma,\tau) &\equiv &\frac {\langle \textbf{x}'(\sigma_{1}) \cdot
 \textbf{x}'(\sigma_{2})\rangle} {\langle \textbf{x}'(0) \cdot
 \textbf{x}'(0)\rangle} \\ \nonumber
 &=& \frac{\langle \textbf{p} (\sigma_{1})  \cdot \textbf{p} (\sigma_{2})  \rangle -
 \langle \textbf{p} (\sigma_{1})  \cdot \textbf{q} (\sigma_{2})  \rangle}{2(1-\langle\dot{\textbf{x}}^2 \rangle)}.
\end{eqnarray} 

\begin{figure}
\includegraphics[width=0.9\columnwidth]{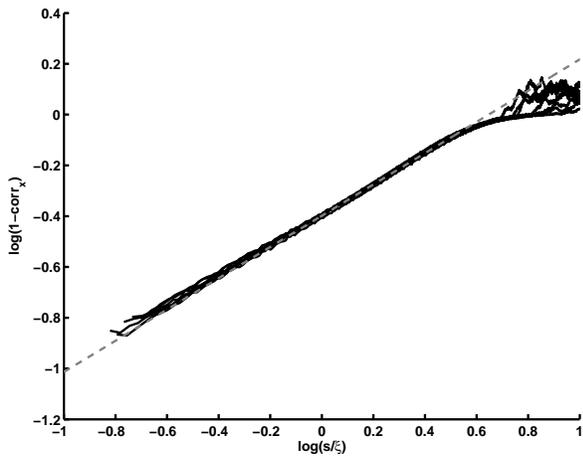}
\caption{\label{radCorrs}Solid black lines indicate 2 point tangent vector correlation functions in the 
radiation era for 7
  equally spaced times throughout the scaling epoch ($\tau\in[64, 128]$) with dashed grey fit calculated 
from an average of all 7 sets of
  parameters fitted up to $\log(s/\xi)=0.5$.}
\end{figure}

\begin{figure}
\includegraphics[width=0.9\columnwidth]{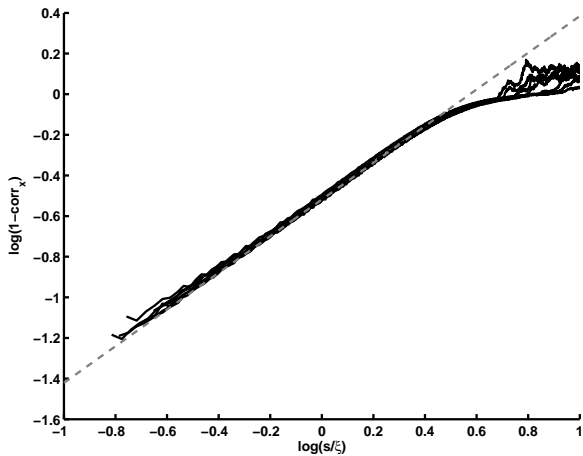}
\caption{\label{mattCorrs}Correlations and fit for the matter era as in
  Fig.\ \ref{radCorrs} with fit performed on curve to $\log(s/\xi)=0.3$. }
\end{figure}

Correlations for opposite moving modes 
$ \langle \textbf{p} (\sigma_{1})  \cdot \textbf{q} (\sigma_{2})  \rangle$
 will be subdominant to the build up of those moving in the same direction along the string 
 $ \langle \textbf{p} (\sigma_{1})  \cdot \textbf{p} (\sigma_{2})  \rangle$. Identifying that $-\textbf{p}(\sigma)
\cdot\textbf{q}(\sigma)=1-2\dot{\textbf{x}} ^2$, averaging over an ensemble of segments and many 
hubble times allows an approximation to first order of $-\langle\textbf{p}(\sigma_1)\cdot\textbf{q}
(\sigma_2)\rangle=1-2\langle\dot{\textbf{x}} ^2\rangle$ for small $s$ giving
 $$2(1-\langle  \dot{\textbf{x}}^2 \rangle)(1-\textrm{corr}_{\textrm{x}}(\sigma, \tau))=1-\langle\textbf{p} 
(\sigma)\cdot \textbf{p}(0)\rangle.$$
The equations of motion can be rewritten in terms of the evolution of  $\langle \textbf{p}(\sigma)\cdot 
\textbf{p}(0) \rangle$.
\begin{equation}
\partial_{\tau} \langle 1-\textbf{p}(\sigma)\cdot \textbf{p}(0) \rangle = - \frac{2\dot{a}}{a}
(1-2\langle\dot{\textbf{x}}^2 \rangle )\langle 1- \textbf{p}(\sigma) \cdot \textbf{p}(0) \rangle
\end{equation}
which integrates to the form
\begin{equation}
\label{functionForPP}
\langle 1-\textbf{p}(\sigma) \cdot \textbf{p}(0) \rangle = f(\sigma)\tau^{-2\nu(1-2\langle
 \dot{\textbf{x}}^2 \rangle )}
\end{equation}
with  $\nu$  defined as scale factor $a\propto\tau^{\nu}$. The correlator must be a function of 
$s/\tau$ if it is to exhibit scaling. Given that $s=\epsilon\sigma$, we need the time dependence of 
of $\epsilon$ 
(Eq.\ \ref{eqn4eps}) $$\frac{\dot{\epsilon}}{\epsilon}=-2\frac{\dot{a}}{a}
 \dot{\textbf{x}}^2 \Rightarrow \epsilon \propto \tau^{-2\nu\langle \dot{\textbf{x}}^2 \rangle}.$$
Hence $$\frac{s}{\tau}=\frac{\epsilon\sigma}{\tau}\propto \sigma\tau^{-1-2\nu\langle \dot{\textbf{x}}^2 \rangle}.$$ Given that Eq.\ (\ref{functionForPP}) has  a power law 
form
 $$ \langle 1-\textbf{p}(\sigma) \cdot \textbf{p}(0)\rangle \propto \Big( \frac{s}{\tau}
 \Big)^{2\chi} $$
we find
\begin{equation}
\label{exponent}
 2\chi = \frac {2\nu (1-2\langle \dot{\textbf{x}}^2 \rangle)} {1+2\nu\langle \dot{\textbf{x}}^2 \rangle}.
\end{equation}
The tangent vector correlator should also scale 
\begin{eqnarray}
\label{powerLaw}
& 1-\textrm{corr}_{\textrm{x}}(\sigma, \tau) &= \textrm{A}\Big(\frac{s}{\xi}\Big)^{2\chi}\\
\label{powerLawFit}
\Rightarrow&\displaystyle{\frac{\partial^2\langle r^2\rangle }{\partial\sigma^2}}
  &= \langle\textbf{x}'\cdot \textbf{x}' \rangle\Big( 1-
  \textrm{A}\Big(\frac{s}{\xi}\Big)^{2\chi}\Big)
\end{eqnarray}
In order to test the prediction, 
the second derivative of the 2-point function $\langle
r^2(s,\tau)\rangle$  is taken numerically to find a fit for the
parameters in this analytic result. A least squares fit is used to
optimise the parameters in the function for $1-\textrm{corr}_{\textrm{x}}$ taking a
nominal standard deviation equivalent to the length $s$ to weight smaller
$s$ appropriately on the logarithmic scale. Noise is reduced by
averaging the result
over 20 simulations and additionally averaging over a block of
$\Delta s=20$ and centring the mean value. It
should be noted that the smoothing process was not found to alter the
parameters or the shape of the $1-\textrm{corr}_{\textrm{x}}$ function, notably at
small $s$, but allowed
the least squares fit to converge more quickly. These functions are
shown for the radiation era Fig.\ \ref{radCorrs} and
matter era Fig.\ \ref{mattCorrs} simulations. The correlations over long distances
reflect smoothness on large scales and a clear power law
demonstrated on smaller scales from a few times the correlation length down to
string width scales. The average parameter values over 7 seven equally
spaced times in the scaling epoch of the simulation have also been calculated for $\nu=0.5$
and $\nu= 1.5$ as an investigation of transitions between
eras. Results are given in Table \ref{results}.
\begin{table}
\begin{tabular}{|c|ccc|c||c|}
\hline
$\nu$ & $\langle \textbf{x}' \cdot \textbf{x}' \rangle$ & $A$ & $2\chi$ &
$\langle \dot{\textbf{x}}^2 \rangle_{\mathrm{2\chi}}$ &
$\langle \dot{\textbf{x}}^2 \rangle_{\mathrm{G}}$ \\
\hline
$0.5$ & $1.0$ & $0.86\,$ & $0.44$ & $0.23$ &  $0.22$\\
$1$ & $0.97$ & $0.40\,$ & $0.62$ & $0.26$ & $0.27$\\
$1.5$ & $0.96$ & $0.85\,$ & $0.71$ & $0.27$ &  $0.27$\\
$2$ & $0.92$ & $0.30\,$ & $0.90$ & $0.27$ & $0.26$ \\
\hline
\end{tabular}
\caption{\label{results}Table to show parameters found from 2 point
  correlation data fitted to predicted power law Eq.\ (\ref{powerLaw}) of
  Ref.\cite{Polchinski:2006ee}. Correlation functions plotted in
  Figs.\ \ref{radCorrs} and  \ref{mattCorrs}. The implied mean square network
  velocity, $\langle \dot{\textbf{x}}^2 \rangle_{\mathrm{2\chi}}$ is calculated from the exponent according 
to
  Eq.\ (\ref{exponent}). These values are compared with the direct
  calculation of network velocities in Fig.\ \ref{velocities} and quoted for $\langle \dot{\textbf{x}}^2 
\rangle_{\mathrm{G}}$, calculated from the field gradients. }
\end{table}
\begin{figure}
\includegraphics[width=0.9\columnwidth]{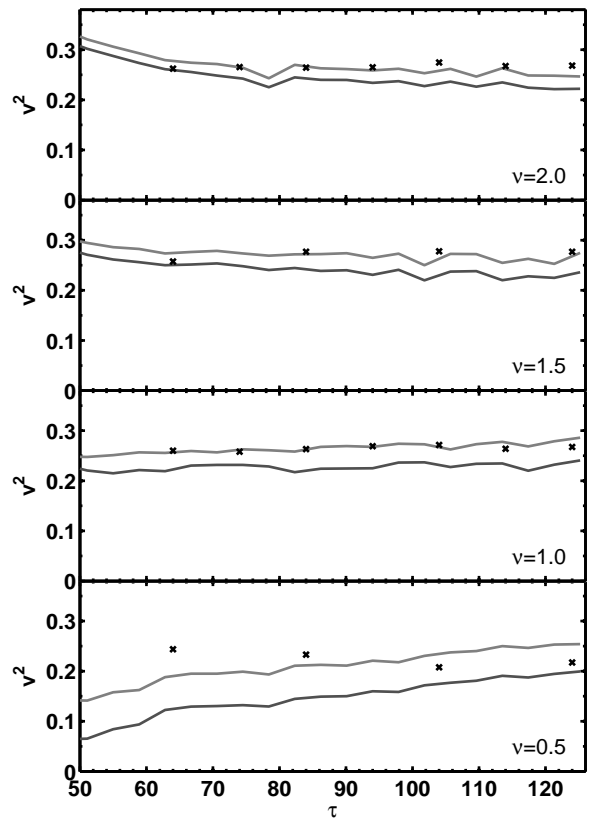}
\caption{\label{velocities}From bottom to top $\langle \dot{\textbf{x}}^2 \rangle$ is
  calculated for $\nu= 0.5, 1, 1.5, 2$. Crosses mark values calculated from
  $2\chi$ exponent according to Eq.\ (\ref{exponent}); dark grey shows $\langle \dot{\textbf{x}}^2 \rangle_
{\mathrm{F}}$ calculated from Eq.\ (\ref{EBvel}); light grey
shows $\langle \dot{\textbf{x}}^2 \rangle_{\mathrm{G}}$  calculated from Eq.\ (\ref{GradVel}). }
\end{figure}
The parameter $2\chi$ provides an estimate for the
average network velocity squared $\langle \dot{\textbf{x}}^2 \rangle$ via
Eq.\ (\ref{exponent}) and
its value can be cross-checked with that calculated
directly from the simulation Fig.\ \ref{velocities}. Two methods are used for {estimating} the average 
network velocity. One {uses} the electric $\bE$ and magnetic $\bB$ components from the field strength 
tensor and the other the {canonical momentum and spatial gradients of the field, $\Pi = \dot\phi $ and $
\bD \phi=\partial\phi/\partial \textbf{x}+ i \bA \phi$.  
Denoting the estimators $\langle \dot{\textbf{x}}^2 \rangle_{\mathrm{F}}$ and $\langle \dot{\textbf{x}}^2 
\rangle_{\mathrm{G}}$, we have 
\begin{equation}
\label{EBvel}
\gamma^2_F\langle \dot{\textbf{x}}^2 \rangle_{\mathrm{F}} =\frac{\bE_{\Lag}^2}{\bB_{\Lag}^2}
\end{equation}
and 
\begin{equation}
\label{GradVel}
 \gamma^2_G\langle \dot{\textbf{x}}^2 \rangle_{\mathrm{G}} = \frac{\Pi^2_{\Lag}}{(\bD\phi)^2_{\Lag}},
\end{equation}
where $\gamma^2_F$ and $\gamma^2_G$ are calculated using $\langle \dot{\textbf{x}}^2 \rangle_
{\mathrm{F}}$ and $\langle \dot{\textbf{x}}^2 \rangle_{\mathrm{G}}$ respectively, and 
the subscript $\Lag$  denotes a Lagrangian weighting of a field $X$
according to 
$${X}_{\Lag} = \frac{\int\textrm{d}^3x \,\,{X}\,\,\mathcal{L}}{\int\textrm{d}^3x \,\,\mathcal{L}}.
$$}
We see that the measured velocities agree well with those inferred from the slope of the correlation 
function, strong evidence that the model of \cite{Polchinski:2006ee}  describes the dynamics of long 
string in the Abelian Higgs model well.

{Finally we note that our root mean square (RMS) velocities are approximately $0.5$ in both the matter and radiation era. 
These are significantly lower than measured in Nambu-Goto simulations, about 0.66 in the radiation era 
and 0.61 in the matter era. This is likely to be a result of backreaction from massive radiation, not 
included in Nambu-Goto simulations.
Our velocities are, however, in agreement with the uncorrected RMS velocities of Ref.\ 
\cite{Moore:2001px}, which are measured in a field theory simulation using the positions of the zeros of the field.}

\subsection{Initial Conditions and Relaxation to Scaling}

It is important to test for any dependence of our results upon the initial conditions chosen and to fully explore the approach to scaling. To achieve these two goals we have performed additional simulations with an initial phase in which the Hubble damping term $2\dot{a}/a$ in Eqs.\ (\ref{e:EOMHiggs}) and (\ref{e:EOMGauge}) is replaced by a constant, which we set to unity. This initial phase of artificial damping continues until a time when the Hubble damping would have been far smaller and hence string velocities are heavily reduced by the time we switch over to normal Hubble damping to complete the simulation. This gives us an alternative initial condition in which the string network is smooth and  slowly moving while radiation is negligible.

The effects on the network length scale $\xi$ for simulations on $512^3$ lattices, seen in  Fig.\ \ref{xi_scaling}, shows the rate of growth of  $\xi$ is heavily retarded during the initial phase with $\xi \propto \tau^{1/2}$ as expected for over-damped motion \cite{Martins:1995tg}. Once the constant damping is switched to Hubble damping, the evolution of $\xi$ quickly transitions to the same growth rate seen in the primary simulations.

\begin{figure}
\includegraphics[width=0.9\columnwidth]{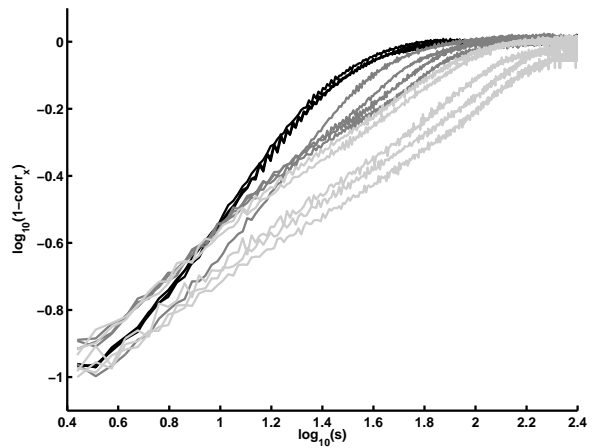}
\caption{\label{damp55}Log of the tangent vector correlation function $1-\textrm{corr}_{\textrm{x}}$ 
against  log of the comoving string coordinate length $s$ for a radiation era simulation with constant
  damping until $\tau=55$ on a $768^3$ lattice. Black curves show the correlation function at times before, at and immediately after damping is released to ordinary Hubble damping.  In dark grey are correlators at times when then the exponent is in transition from the constantly damped evolution but before scaling is achieved  ($\tau = 62, 74, 80, 88$). Scaling is re-established for the correlators plotted in light 
grey for times $\tau>96$.}
\end{figure}

\begin{figure}
\includegraphics[width=0.9\columnwidth]{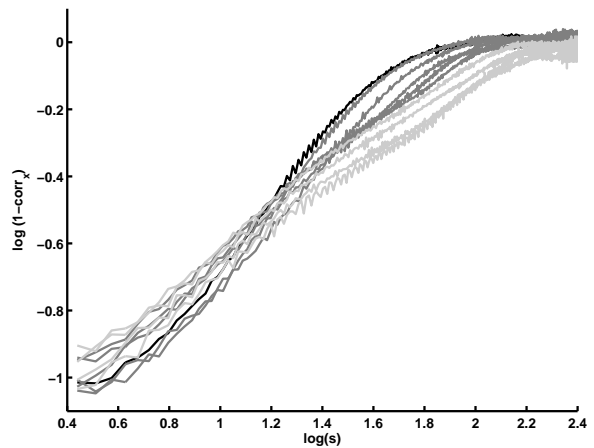}
\caption{\label{damp100}Radiation era simulation on $768^3$ lattice with constant
  damping until $\tau=100$. The black curve shows log($1-\textrm{corr}_{\textrm{x}}$)  at $
\tau=96$;  before artificial damping is switched off.  In dark grey are correlators shown at times   $\tau = 104, 112, 120, 128, 136$ when then the exponent is in transition. Scaling correlators are shown in light grey for $\tau=144, 160, 176, 192$. Noise is greater for these curves compared with Figs.\ \ref{radCorrs} and \ref
{mattCorrs} as data
 is taken from just one simulation.}
\end{figure}

The effects on the correlation functions are shown in Figs. \ref{damp55} and \ref{damp100}. For clarity, the log of $1-\textrm{corr}_{\textrm{x}}$ is plotted against the log of the comoving length along the string $s$, which separates the curves so the changes in gradient are more apparent.
It is clear from the tangent vector correlations that when the 
evolution is artificially damped, the small scale structure is quite different. The initial slope for $\log(s)<1.4$ is 
close to unity, as expected for smooth strings, which is maintained for a short time.
Then $1-\textrm{corr}_{\textrm{x}}$ begins a transition towards a new exponent as scaling is attained for 
the new undamped (Hubble damping only) evolution. The power law changes first at small scales and moves up to larger 
scales as the evolution adjusts to the new conditions. Once the exponents have relaxed, the small scale 
structure behaves and scales just as in an undamped simulation with no recollection of the initial 
damping conditions.

\begin{figure}
\includegraphics[width=0.9\columnwidth]{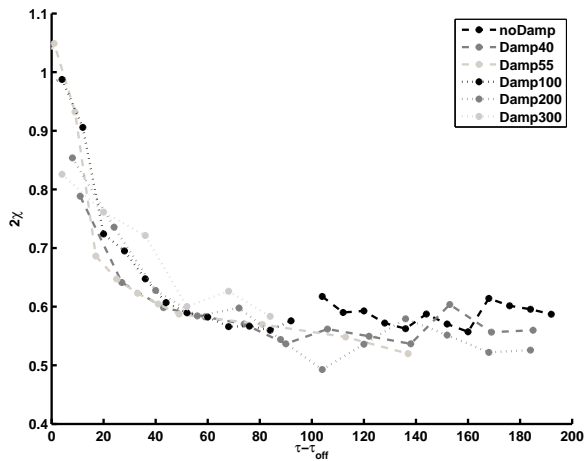}
\caption{\label{dampRelax} Relaxation of the parameter $2\chi$ after constant damping is switched off in a 
number of $768^3$ radiation era simulations with $\tau_{\mathrm{off}}=0, 40, 55, 100, 200, 300$. }
\end{figure}

The exponents are found by setting $\langle \textbf{x}' \cdot \textbf{x}' \rangle=0.97$ as found for the 
radiation era simulations on a $512^3$ lattice which are also consistent with that found for a single 
$768^3$ simulation. So in this case only a 2 parameter fit is made for Eq.\ (\ref{powerLawFit}); $2\chi$ 
and $A$. $A$ is found to be constant and consistent in both a damped and undamped evolution so we 
discuss the relaxation time scale in terms of the time taken for the gradient parameter, $2\chi$, to 
transition back into scaling. There is little effect on the value of $2\chi$ once  the new scaling value is 
reached after damping and becomes approximately 0.6, the same as found during the scaling epoch for 
a completely undamped simulation. The profile of this relaxation is compared for different damped 
simulations and shown in Fig.\ \ref{dampRelax}. 

Although not conclusive, there is evidence that the relaxation time, from the moment damping is turned 
off $\tau_{\mathrm{off}}$, is approximately $\xi$. {One can see the trend most clearly in the value of $
\tau-\tau_{\rm off}$ at $2\chi \simeq 0.7$.  This is explicable if the transition is triggered by 
intercommutation and the generation of kinks, which would have the observed effect of reducing the 
correlation at small scales.}

Figs.\ \ref{damp55} and \ref{damp100} show the correlators settling into a scaling regime behaving 
much as the
evolution with Hubble damping only. We recall also that the network length scale $\xi$ returns to the 
expected linear behaviour, Fig.\ \ref{xi_scaling}, over perhaps an even shorter time frame. We interpret 
these features as good evidence that calculations can  be trusted to not depend on initial conditions.

\subsection{Comoving String Core Width}

\begin{figure}
\includegraphics[width=0.9\columnwidth]{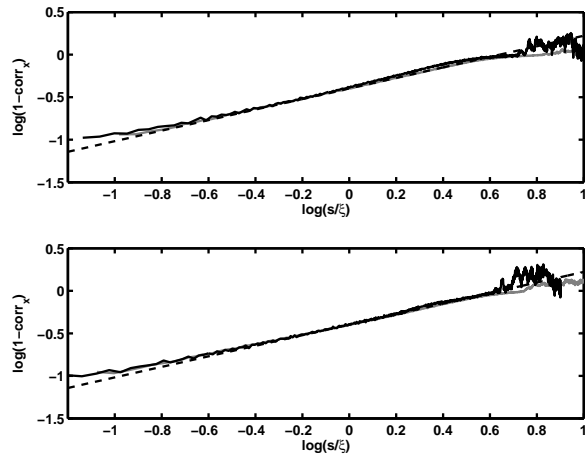}
\caption{\label{sTest}Grey lines show correlations for $\kappa=0$ and black for
  $\kappa=1$. In the top figure the comparison is made at time $\tau= 85$ and at the
  bottom is the comparison at time $\tau=115$.}
\end{figure}

Simulations conducted in the radiation era are possible with the true
equations of motion using $\kappa=1$ in Eqs.\ (\ref{e:EOMHiggs})-(\ref{e:EOMGauge}) and
correlation results
for an average of $5$ simulations
for the correlation function results are tested against $\kappa=0$ results. Fig.\ \ref{sTest} shows a 
comparison of the correlations at two different times in the simulation for both the $\kappa=0$ and the
$\kappa=1$ cases. The difference in the results is surprisingly insignificant for this
calculation and no correction is felt necessary.

\subsection{Testing Gaussianity}

The 4-point correlation functions $\langle r^2(s,\tau)\rangle^2$
and $\langle (r^2(s,\tau))^2\rangle$  are also calculated to test
for gaussianity.

Denoting each spatial component of $r$ as $\Delta_i \equiv
\textbf{x}_i(\sigma_1)-\textbf{x}_i(\sigma_2)$, with $\sigma_1$ some initial
base point for the measurement, the fourth moments formula is written
\begin{eqnarray}
\langle (r^2(s,\tau))^2\rangle& = &\langle (\Delta \cdot
\Delta)^2\rangle \nonumber \\ &=& \delta_{ij}\delta_{jk}
(\langle \Delta_i \Delta_j\rangle\langle \Delta_k
\Delta_l\rangle \nonumber \\&&+\langle \Delta_i \Delta_k\rangle\langle \Delta_j
\Delta_l\rangle\nonumber  \\&&+\langle \Delta_i \Delta_l\rangle\langle \Delta_j
\Delta_k\rangle  )
\end{eqnarray}
Then gaussianity would allow contraction on all pairs from Wick's Theorem
so that $\langle (\Delta_i \Delta_j)\rangle =
\frac{1}{3}\delta_{ij}\langle \Delta^2\rangle$ and the ratio of the 4
point functions should
behave as
$$\langle (r^2(s,\tau))^2\rangle = \frac{5}{3}\langle
r^2(s,\tau) \rangle^2$$

\begin{figure}
\includegraphics[width=0.9\columnwidth]{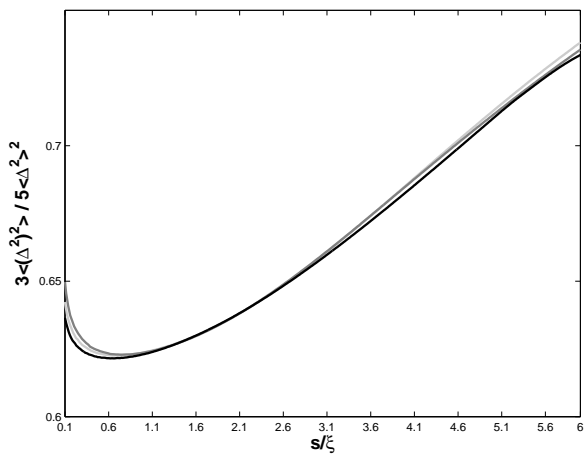}
\caption{\label{4pointRad}The ratio of the 4-point correlation functions 
to the Gaussian prediction in the
  radiation dominated era is shown against the dimensionless distance parameter
  $s/\xi$. A Gaussian 4-point correlator should give a value 1: Gaussianity is therefore not evident at any scale.}
\end{figure}

Fig.\ \ref{4pointRad} shows for the radiation era  that on all scales the ratio of the 4 point
correlations is not constant and not $5/3$. The 4-point correlators in the matter era 
(not plotted) behave in a similar way.

In the model for loop production proposed by Polchinski and collaborators  
\cite{Polchinski:2006ee,Dubath:2007mf, Polchinski:2007rg},  it is argued that non-gaussianity 
does not affect the power laws derived for the 2-point correlation function or the loop production function.  
We have confirmed that the 2-point correlation function is in accord with their model, 
but in the next section we will see that the loop production function is not. 
If we accept the arguments of Polchinski et al, which are based on scaling, 
another reason must be found to account for the difference.

\section{Loop Distributions}
\label{loopDistributions}

Loop production in cosmic string networks is still 
a subject of some debate 
despite  a number of recent numerical investigations using the Nambu-Goto 
approximation
\cite{Ringeval:2005kr, Olum:2006ix, martins:043515}, 
taking advantage of improvements in computational facilities and algorithms,
and focusing on small
scale structure and loop production rates. 

The crucial quantities in question are the loop (length) distribution function and the loop production function. Unfortunately, 
the different groups measure different quantities, and emphasise different features, so the results are difficult to compare.

Those that measure the loop production function \cite{Olum:2006ix, martins:043515}
find that it peaks at a small scale, with a power-law rise \cite{Olum:2006ix}, 
and a less prominent feature at about a tenth of the horizon length $t$.
The identity of the small scale is not clear, but on inspection of the data \cite{martins:043515,Olum:2006ix}, it appears to be related to (and at least no greater than) the initial comoving correlation length.  
Full scaling requires that the only scale in the distribution and production functions should be $t$: 
the peak therefore does not scale. Furthermore, it is found that the amplitude of the power law does not scale either \cite{Olum:2006ix}.

Measurements of the loop distribution function on the other hand  \cite{Ringeval:2005kr}, show a peak at the initial numerical cut-off, and scaling at intermediate scales.  The peak is understandable as a transient from the initial evolution, but as the distribution function is essentially the time integral of the production function, the intermediate range scaling is a puzzle.

It has been suggested that the non-scaling of the loop production function is a  transient effect \cite{martins:043515,Olum:2006ix}, and that the peak should eventually disappear altogether \cite{Olum:2006ix} or start scaling if only a large enough simulation could be performed \cite{martins:043515}. 
However, the evidence that a power-law with a small-scale cutoff is a real feature of Nambu-Goto string networks has been strengthened thanks to the agreement with Polchinski and collaborators' 
model of loop production \cite{Polchinski:2006ee,Polchinski:2007rg,Dubath:2007mf}. There is also no evidence for scaling of the peak in the loop production function from visual inspection of the graphs in Refs.\ \cite{martins:043515,Olum:2006ix}.

Accepting the evidence of a power-law form for the loop production function, a small-scale cut-off  is required to keep the total energy loss finite.  The conventional string scenario demands full scaling, and invokes gravitational radiation reaction 
to change the loop production scale to a   
constant fraction of the horizon size, which is $(G\mu)^{1+ 2\chi}t$,  according to Ref.\ \cite{Polchinski:2007rg}. However, there are no network simulations including gravitational radiation reaction so this is still a conjecture.  It could equally well be that loop production really does not scale as the Nambu-Goto simulations suggest; this does not prevent the energy density of the long string network from scaling. Furthermore, if the small-scale cut-off is the string width \cite{Vincent:1996rb}, it is necessary to perform field theory simulations in order to include the true small-scale physics.  

Previous field theory simulations 
\cite{PhysRevLett.80.2277,Moore:2001px}
have not studied the loop distributions in any detail, but it is already clear that their properties are very different from the Nambu-Goto versions.
The number
of loops in the simulation volume is substantially less, which prompted the suggestion \cite{PhysRevLett.80.2277} 
that the network could lose energy to classical radiation directly rather than via the 
production and eventual decay of loops. Arguing in favour of loop production, it was pointed out in \cite{Moore:2001px} 
that even if all the energy is lost to ``core" or ``proto"-loops (loops whose length is of order the 
string width) that the number density would be very low anyway, approximately 
$t^{-3}$.  It was also conjectured that these core loops would eventually grow if a large enough simulation could be performed.

In the first part of this section we test the hypothesis that a substantial fraction of the energy loss from long strings is in the form of core loops.
The impression given by visualisations such as  
Fig.\ \ref{radiating} is that direct radiation appears to be very important, 
although it is very difficult to tell the difference between a large amplitude excursion 
by the Higgs field and a core loop. However, if energy loss into core loops is important we would expect to find core loops near long strings, and our first test is to look for these correlations.

\begin{figure}
\includegraphics[width=\columnwidth]{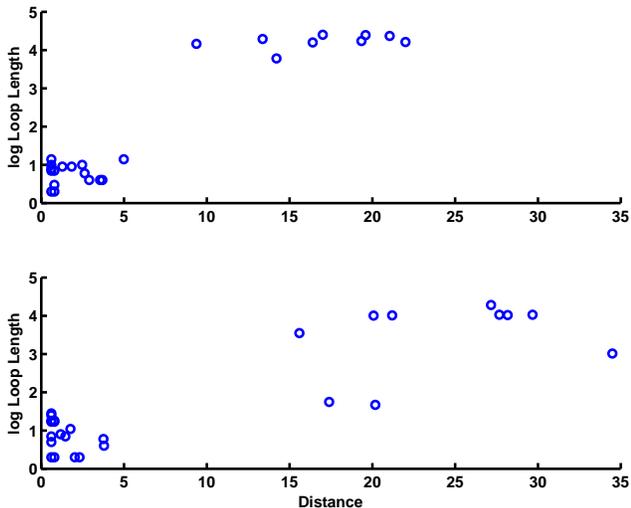}
\caption{\label{protoDist} Distance of core loops to other string in the radiation era in the top plot and 
matter era in the lower plot. Small loops lie close together in clusters in the voids between long strings at 
about half the average interstring distance, $\sim \xi/2$. Emission of core loops directly from long string 
is not detected.  
}
\end{figure}

We define a core loop as a loop with the minimum number of lattice segments to create a closed loop 
(i.e. 4 linked segments, located as described in Sec.\ \ref{lengthMeasures}). With $\Delta x=0.5$, the  core loop length is approximately the string width.  We then measure the distance from core loops to the closest point on a neighbouring piece of string and the length of the string to which it is closest. The results are shown in Fig.\ \ref{protoDist}. Interestingly, it is seen that core loops lie close 
to other very small loops and that these clusters or isolated core loops lie at distances of order  half the 
correlation length from long string; between two long strings. As loops collapse they appear to fragment 
into clusters of very small loops but the lack of core loops close to long strings argues against small loop 
production being a significant 
source of energy loss from long strings.

\begin{figure}
\includegraphics[width=0.9\columnwidth]{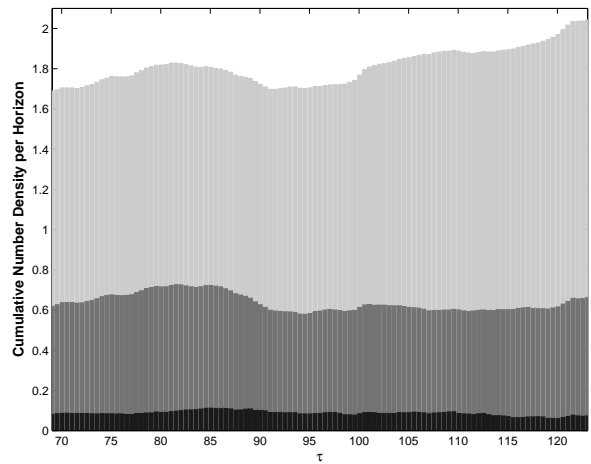}
\caption{\label{loopDistn}Number of loops per horizon volume in the radiation era over the scaling
  epoch $\tau \in[64,128]$ with data smoothed by averaging results over blocks of $10\tau$. Dark grey 
represents proportion by number of `proto'
  loops of size the order of the string core width, $\delta$; mid-grey
  represents loops up to the size of the correlation length, $\xi$ at that
  time in the simulation; light grey represents all string that is
  greater in length than $\xi$ and considered to be infinite string in
  this case.}
\end{figure}

We can also check the hypothesis by looking at the number distribution of loops in field theory simulations, using a large number of runs. 
The length scales of interest are the string width (core loops)
and the network  
correlation length, $\xi$, defined as Eq.\ (\ref{xiDef}).  Fig.\ \ref{loopDistn} shows cumulatively the number 
densities of loops per horizon in the radiation era over the 
conformal time range $64<\tau<128$ when the network is scaling. The loops are divided into those of 
length 4 links (core loops), those up to length $\xi$, and those beyond.  
Core loops are seen to occupy a constant small fraction of the number density.  In each of these classes 
the number density of loops per horizon appears to be constant. The number of core loops per horizon is 
very small, of order $0.1$.

We can estimate whether this is consistent with core loops being a significant channel of energy loss for 
long strings. If a network with comoving length scale $\xi = x\tau$ decays into loops of size $\bar l$, then their lifetime should also be $\bar l$, given the shrinking mechanism outlined in Section \ref{CSD}.
The number of loops per horizon volume $\tau^3$ should then be $\sim 1/x^2$.  Given that $x^2 \sim 0.1$, there are roughly 100 
times too few core loops if they were to take a significant amount of energy away from long strings.  

Our result seems to be in contradiction to Ref.\ \cite{Moore:2001px}, who use a fit to the Velocity-dependent One-Scale 
model \cite{PhysRevD.54.2535} to argue that loop production is significant in their field theory 
simulations. However, they do not give absolute values of the loop distribution function and so it is not 
possible to compare the results directly.  One possible resolution, explored in more detail below, is that horizon-size loops with lifetime $\sim t$ are carrying away an appreciable fraction of the energy.

To study the loop distribution and production functions in more detail, we must model both the
production and shrinking of loops. We denote the loop distribution function in terms of the  
cosmic time $t$ and physical length $l_p$ as $n_p(t, l_p)$, where $l_p=a(t)l$ and $l$ is the comoving loop length, which is given in terms of the string variables $\epsilon$ and $\sigma$ by  $l=\int \epsilon \mathrm{d}\sigma$. 
We denote the comoving loop distribution function in conformal time as $n(\tau, l)$.  Then the number 
density of loops $n_p(t,l_p)\mathrm{d}l_p$ in physical length interval $[l_p, l_p + \mathrm{d}l_p]$ is 
related to the comoving number density of loops $n(\tau, l)\mathrm{d}l$ in interval $[l, l + \mathrm{d}l]$  
by
\begin{eqnarray}
&n(\tau, l)\mathrm{d}l &= a^3 n_p(t,l_p)\mathrm{d}l_p\nonumber\\
\label{physComoveN}
\Rightarrow&n(\tau, l)&= a^4 n_p(t,l_p)
\end{eqnarray}

The equation governing the loop distribution function is
\begin{equation}
\frac{\partial n_p}{\partial t}+3H n_p +\frac{\partial l_p}{\partial t}\frac{\partial  n_p}{\partial l_p} = P_p
(t,l_p)
\end{equation}
where we introduce the loop production function in physical units $P_p$. We also take into account 
energy loss from loops which we shall assume takes place at a constant rate
 such that $\textstyle\frac{\partial l_p}{\partial t}=-\lambda$. We estimate $\lambda \sim 
\mathcal{O}(1)$ from the properties of the energy loss mechanism outlined in 
Sec.\ \ref{CSD}\footnote{\mbox{Gravitational radiation would give $\lambda \sim \Gamma G\mu$ were it to be included}}. 
Using Eq.\ (\ref{physComoveN}) we can relate the comoving number 
density distribution and the loop production function in comoving units
\begin{eqnarray}
\label{loopProdFn}
 \frac{\partial n}{\partial \tau}-\frac{\dot{a}}{a}n  -\lambda \frac{\partial  n}{\partial l} =P(\tau,l)
\end{eqnarray}
where $P(\tau,l)=a^5P_p(t,l_p)$.

\begin{figure}
\includegraphics[width=0.9\columnwidth]{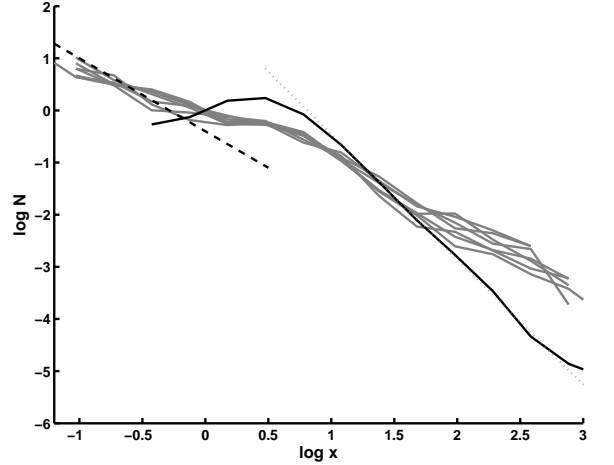}
\caption{\label{scalingLoops}  The loop number density $N=\tau^4\Delta n/\Delta l$, per
  comoving logarithmic bin length $\Delta l$ is shown for 7 equally spaced times throughout the
  scaling epoch of the radiation era against the dimensionless ratio $x=l/\xi$. This
  is compared with the very early time $\tau=10$ case in solid black which is
  compatible for $l \gg \tau$ with slope -5/2
  (shown dotted) as predicted in \cite{Vachaspati:1984dz}. The slope predicted by the model of Ref.\ \cite
{Dubath:2007mf} including radiative effects at small scales as in Eq.\ (\ref{NpowerIncRad}) is shown with 
$\alpha+1=-1.4$ (dashed black), where $\alpha$ is the slope of the loop production function. }
\end{figure}

\begin{figure}
\includegraphics[width=0.9\columnwidth]{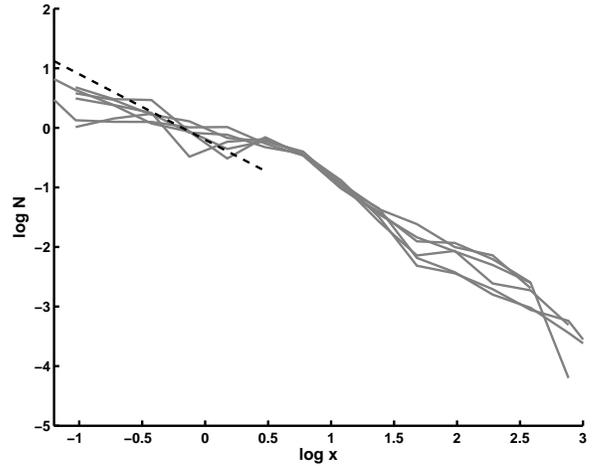}
\caption{\label{loopsMatter}Figure showing number density of loops
  during the scaling epoch in the matter era with slope $\alpha+1=-1.1$ predicted by the model of Ref.\ \cite{Dubath:2007mf} including radiative effects at small scales as in Eq.\ (\ref{NpowerIncRad}) shown in dashed black. 
  }
\end{figure}

 Assuming scaling, the comoving loop production function and number density distribution behave as
\cite{vilenkin} 
$$n(\tau, l)=\textstyle{\frac{1}{\tau^4}N(x)}\,\, \mathrm{  and  } \,\,P(\tau, l)=\textstyle{\frac{1}
{\tau^5}f(x)}$$ 
for functions $N$ and $f$ of the dimensionless ratio of loop length to horizon size $x=l/
\tau$.  Rewriting Eq.\ (\ref{loopProdFn}) in terms of $N$ and $f$  one obtains (with $a\propto\tau^{\nu}$),
 $$(x  +\lambda) N'(x) +(\nu+4)N(x) = -f(x)$$
 with solution
 \begin{eqnarray}
\label{NdensWithRad}
N(x)= (x+\lambda)^{-(\nu+4)}\int^\infty_x f(x') (x'+\lambda)^{\nu+3}\,\,dx'.
\end{eqnarray}
Numerical simulations suggest a power law for loop production, $f\propto x^{\alpha}$ below $x\sim 1$. If radiative 
effects can be neglected ($x\gg \lambda$), and making the reasonable assumption that  $f$ vanishes for $x \gg 1$, we have from Eq.\ (\ref{NdensWithRad}) 
 \begin{equation}
 \label{NpowerNoRad}
N\propto f \propto x^{\alpha}.
 \end{equation}
For length scales where  radiative effects are strong ($x\ll \lambda $)
 \begin{equation}
 \label{NpowerIncRad}
 N\propto x^{\alpha+1}.
 \end{equation}
 To make our measurement we define the comoving loop number density in a length interval $\Delta l = l
$ $$\Delta n = \int^{2l}_{l} n(\tau, l')\mathrm{d}l'.$$ Figs.\ \ref{scalingLoops} and \ref{loopsMatter} show 
an 
estimate for $$N(x)=\tau^4 \displaystyle\frac{\Delta n}{\Delta l}$$ taken from the average of $20$ runs in 
radiation and matter eras respectively. 
 The solid black line in Fig.\ \ref{scalingLoops} the the initial loop distribution function which shows 
good agreement with the expected power law of slope $-5/2$, \cite{Vachaspati:1984dz}.

For a network that has reached scaling, 
the analytic model of Ref.\ \cite{Polchinski:2006ee}, further refined
by Dubath et al in  \cite{Dubath:2007mf}, proposes that loop density
is dominated by recently
produced loops.
They derive a function for loop production which they extrapolate to loop number density distribution 
under the assumption that radiative effects can be ignored and obtain
\begin{equation}
\label{modelNdens}
f\propto \Big(\frac{l}{\tau}\Big)^{2\chi-3} 
\end{equation}
with $\chi$ defined in Eq.\ (\ref{exponent}). If radiative effects are considered the exponent  for the 
number density distribution function will be higher by $+1$ at small length scales by the arguments from 
Eqs.\ (\ref{NdensWithRad}) and (\ref{NpowerIncRad}). This is also consistent with the analytic findings of Rocha in Ref.\ \cite{Rocha:2007ni} where the loop distribution model is enhanced by including of the effects of gravitational radiation.
From the values of $2\chi$ calculated from our simulation results and given  in 
Table \ref{results}, this  would give exponents of $\alpha_\mathrm{R} +1\sim -1.4$ in the radiation 
era and $\alpha_\mathrm{M} +1\sim-1.1$ in the matter era which are compared to our data in Figs.\ \ref
{scalingLoops} and \ref{loopsMatter}. The agreement remains less than convincing, particularly for the radiation era.

Olum et al
\cite{Olum:2006ix} calculate the loop production function $f(x)$ from their Nambu-Goto simulations.
The function drops from a small-scale peak with a power law consistent with the 
model proposed by Dubath et al \cite{Dubath:2007mf}: $f\propto x^{2\chi-3}$. The  
exponent is calculated using Eq.\ (\ref{exponent}) with velocities taken from simulations of Ref.\ 
\cite{Moore:2001px}; { $v_R = 0.63$ and $v_M=0.57$}. No fit values for the gradient of $\log f$ are quoted in Ref.\ 
\cite{Olum:2006ix} but
pictures showing average gradients of their loop production function in the matter and radiation era  are 
used in  Ref.\ \cite{Dubath:2007mf} to demonstrate their model. Exponents are listed in Table \ref
{resultsCompared}.

Number densities can also be compared with Nambu-Goto simulations of Ref.\ \cite{Ringeval:2005kr} 
who quote a length distribution,
\begin{eqnarray}
\label{NdensWithS}
xN(x)\propto x^{p}
\end{eqnarray}
They find a consistent power law over the whole range of their Nambu-Goto simulation with exponents 
$p=-1.6$ for the radiation dominated era and $p=-1.4$ in the matter era.  Given that there is no decay 
mechanism in Nambu-Goto simulations we can infer slopes for the loop production function of $
\alpha=-2.6$ and $\alpha=-2.4$ for radiation and matter eras respectively, in good agreement with the values predicted by the model \cite{Dubath:2007mf} for Nambu-Goto strings. However, even if radiative effects are taken into consideration these results are steeper than 
those predicted for our Abelian Higgs strings, (see Table \ref
{resultsCompared}), which are already too steep to fit our data.

The flatter power law for loop distributions found in the field theory examination Figs.\ \ref{scalingLoops} 
and 
\ref{loopsMatter} can be better explained by invoking loop production and
 fragmentation at horizon scales combined with energy loss at small scales. From Ref.\ 
 \cite{Scherrer:1989ha} it is predicted that unusual power laws for the production of loops can be explained by loop 
fragmentation probabilities, $q$, with $f\propto l^{-2q}$. For small loops where radiative effects are 
strong, number density distribution functions will take the form $N\propto x^{-2q+1}$. Then the steepest 
slope possible would be $-1$ when the fragmentation probability is of order unity.

We also see evidence of loop production at the horizon scale in the small feature visible in the loop 
distribution function at $\log x \simeq 0.5$, which has some similarity to the Nambu-Goto simulations of 
\cite{Olum:2006ix}. It is straightforward to check that the production of one such loop per horizon volume per Hubble time is sufficient to remove a significant fraction of the energy in long strings. Given that these loops are losing length at a rate of order 1 as well as fragmenting, this is consistent with our observation of order one loop per horizon volume at any time. The correlation of core loops with other small loops shown Fig.\ \ref{protoDist} can also be explained by loop 
fragmentation.

Finally, we note that the large $x=l/\tau$ behaviour in our number density analysis remains a puzzle, as it departs from the 
-2.5 
slope expected outside the horizon.  It maybe that the loop distribution 
at these much longer length scales is 
quite sensitive to the finite volume of the simulation \cite{Austin:1993wf}. 
\begin{table}
\begin{tabular}{|l|cc|}
\hline
& $\alpha_R$& $\alpha_M$ \\
\hline
Abelian Higgs Prediction & $-2.4$ & $-2.1$ \\
Nambu-Goto Prediction   & $-2.8$ & $-2.5$ \\
Nambu-Goto Measurement& $-2.6$ & $-2.4$\\
\hline
\end{tabular}
\caption{\label{resultsCompared}Table to show comparison of simulation
  results
  for the exponent $\alpha$ obtained for a power law model for loop production as
  Eq.\ (\ref{modelNdens}) in both radiation (R) and matter (M) eras. The first line shows our values 
calculated using the velocities $\langle \dot{\textbf{x}}^2 \rangle_{2\chi}$ from Table \ref{results}. The 
second line shows the exponent predictions calculated using velocities obtained from  Ref.\ 
\cite{martins:043515} which fits well to the Olum et al Nambu-Goto
simulation  \cite{Olum:2006ix}. The last line shows exponents derived from measured length 
distributions in Nambu-Goto simulations by Ringeval et al
\cite{Ringeval:2005kr}.}
\end{table}


\section{Conclusions and Remarks}
\label{summary}

In this paper we have examined small scale structure, energy loss and loop production 
in cosmic strings found in the Abelian Higgs model (AH), using numerical simulations, and compared 
the results to those from Nambu-Goto (NG) simulations in the literature.  We have also investigated the energy loss mechanism which causes Abelian Higgs strings to scale without gravitational radiation.

We first studied the two
point tangent vector correlation function (Eq.\ \ref{corrX}), for which 
there is a clear prediction based on the NG equations by Polchinski and Rocha (PR)
\cite{Polchinski:2006ee}. 
Their model predicts that the correlator scales, i.e.\ that it is a function of the ratio of distance to the horizon scale, and that this function has a power law form with an exponent determined by 
the mean  square velocity of the string. The correlator is seen to scale very well, and good  agreement with the power law is found right down to the string core
width scale. The mean square velocity is about $0.26$ in both the radiation and matter eras, markedly 
lower than in NG simulations ($0.44$ and $0.37$ and respectively 
\cite{Bennett:1989yp,martins:043515}).  This is likely to be 
due to back-reaction from the massive radiation.

We test the sensitivity of the tangent vector correlator to the initial conditions by damping the strings for an initial interval of various duration. We see that the network regains its scaling form after a time of approximately $\xi$. In this relaxation period we see evidence that the build up of small scale structure appears to move from small scales to large.
{It is plausible that the trigger for the production of small scale structure and the relaxation of scaling is 
intercommutation of long strings.}
The fact that the structure appears bottom-up argues against small-scale structure being due to residual power from the Brownian random walk at 
super-horizon scales, as argued in \cite{Polchinski:2006ee}.

The small scale structure seen in the tangent vector correlations can help to explain the scale separation 
problem described in Sec.\ \ref{CSD}, which is to understand how massive radiation of frequency around about the string width is produced from fields which are apparently changing with a frequency of about the Hubble rate. The solution is that if strings are not smooth on small scales, the radius of 
curvature at these scales is much smaller and frequency of modes of oscillations of the string much higher than the Hubble rate. 

The PR model also has key predictions for loop production \cite{Polchinski:2006ee,Dubath:2007mf}, and shows how 
the small-scale structure accounts for the production of loops at the small-scale cutoff in Nambu-Goto (NG) simulations. We were therefore led to investigate loop distributions in AH simulations also, and to try to connect small scale structure and energy loss in our field theory simulations by looking for core (string width sized) loops. 

We found that the model for loop production in  Ref.\ \cite{Dubath:2007mf} does not describe loop
 distributions in AH simulations, disagreeing with the observed power law. 
Furthermore, a lack of spatial correlation between core loops and long string, and a very low 
overall number density, roughly 0.1 per horizon volume, point towards 
there being no significant energy loss through direct 
production of core loops from long string.

Our results support the statement that significant energy is lost
directly from AH strings.  This can be seen in snapshots from field
theory simulations (Fig.\ \ref{radiating}) with long string radiating energy directly into 
oscillations of the field as Higgs and gauge radiation.  
Our results also indicate that larger loops share this rapid energy-loss mechanism, which 
therefore needs to be added to the   
the standard model for loop distributions. 
We make a simple modification, that  
energy is radiated from loops at a constant rate. 
When coupled with an alternative model of loop production \cite{Scherrer:1989ha}, 
which proposes fragmentation of horizon-sized loops, 
we are able to explain our results for the loop distribution. 

As an aside, a feature of the calculation for the loop distribution function in Ref.\ \cite{Dubath:2007mf} is that the string correlators are assumed to be Gaussian, although it is argued that this is not crucial to the form of the result. The four point correlation in our simulations is explicitly non-Gaussian, but does scale.

So when gravity can be neglected a general picture for the evolution of cosmic string networks is developing. A dominant energy loss 
mechanism works at the small-scale cutoff, which is loop production 
in NG simulations and classical radiation 
in the field theory. Energy is transported from the horizon scale by small-scale structure. 
{In both kinds of simulation loops are produced at the horizon scale,} which then 
fragment with high probability. There is evidence of a population of non-intersecting loops remaining 
stable in NG \cite{Olum:2006ix}, but in AH, where massive radiation is available, loops 
radiate and shrink. 

We note that the small loops in NG simulations behave like particles with mass $\sim \mu\xi_i$, which have similar average energy-momentum properties to classical massive radiation. Thus the striking visual dissimilarity of NG and AH simulations is perhaps misleading: if one interprets the small loops as representing massive radiation and concentrates on the long string network, the differences are greatly reduced. It is interesting to speculate that the small loops might really be massive states for fundamental cosmic strings.

It remains a bit of a puzzle why AH strings have a 
highly suppressed production of small loops.  The PR prediction, based on the NG 
equations of motion, is that small loop 
production should be dominated by the smallest physical scale, which for AH simulations is the 
string width,  but few are observed. As the NG equations of motion break 
down at the string width scale, it is perhaps not surprising to see the break-down of a prediction based on those equations.

In the traditional cosmic string scenario it is assumed that gravitational radiation back reaction plays a 
vital role in smoothing long strings, disguising the small-scale physics of the strings, and setting the scale for loop production. In this scenario most of the energy ends up as long-wavelength gravitational radiation, as opposed to ultra-high energy particles \cite{Vincent:1996rb,PhysRevLett.80.2277}.

It is by no means obvious that gravitational radiation reaction is strong enough to shut off the efficient and rapidly established energy loss mechanism afforded by massive radiation in a field theory. Our simulations show that it is not necessary to postulate gravitational radiation reaction to achieve scaling and a consistent picture of string network evolution. 
Nonetheless, it is still possible that gravitational radiation could operate as traditionally supposed. 
As a  first step, we intend to test the impact of massless radiation on string 
network evolution with an investigation of small scale structure and loop distributions in field theories 
that have massless degrees of freedom, for instance global strings, and the Abelian Higgs model coupled to a dilaton.  

\acknowledgements
Simulations were performed using the COSMOS Altix 3700 supercomputer (supported by SGI/Intel, HEFCE, and STFC) and the University of Sussex HPC Archimedes cluster (supported by HEFCE). We acknowledge financial support from STFC. We thank Ed Copeland, Tom Kibble, Carlos Martins, Ken Olum, Christophe Ringeval, Mairi Sakellariadou and Tanmay Vaspachi for useful discussions.

\bibliography{ref_cosmic_strings}

\end{document}